\documentclass[10pt, conference, letterpaper]{IEEEtran}

\makeatletter
\def\ps@headings{%
\def\@oddhead{\mbox{}\scriptsize\rightmark \hfil \thepage}%
\def\@evenhead{\scriptsize\thepage \hfil \leftmark\mbox{}}%
\def\@oddfoot{}%
\def\@evenfoot{}}
\makeatother 

\IEEEoverridecommandlockouts

\usepackage{url} 
\usepackage{subfigure}
\usepackage{amsfonts}
\usepackage[dvips]{graphicx}
\usepackage{times}
\usepackage{cite}
\usepackage{amsmath}
\usepackage{array}
\usepackage{amssymb}

\newcommand{\bs}{\boldsymbol}

\usepackage{stfloats}
\usepackage{graphicx}
\usepackage{footnote}
\usepackage{booktabs}
\usepackage{amsthm}
\usepackage{array}
\usepackage{algorithmic}
\usepackage{subeqnarray}
\usepackage{cases}
\usepackage{threeparttable}
\usepackage{color}
\makeatletter
\newif\if@restonecol
\makeatother

\usepackage[ruled,vlined]{algorithm2e}
\usepackage{mdframed}

\newtheorem{theorem}{Theorem}

\newtheorem{proposition}{Proposition}
\newtheorem{lemma}{Lemma}

\newtheorem{assumption}{Assumption}

\newtheorem{example}{Example}
\newtheorem{question}{Question}

\ifodd 0
\newcommand{\rev}[1]{{\color{blue}#1}}

\newcommand{\com}[1]{\textbf{\color{red} (Comment: #1) }}
\newcommand{\comg}[1]{\textbf{\color{blue} (COMMENT: #1)}}
\newcommand{\response}[1]{\textbf{\color{blue} (RESPONSE: #1)}}
\else
\newcommand{\rev}[1]{#1}

\newcommand{\com}[1]{}
\newcommand{\comg}[1]{}
\newcommand{\response}[1]{}
\fi

\begin{document}

\title{How to Price Fresh Data\vspace{-5pt}}
\author{
Meng Zhang,
Ahmed Arafa,
Jianwei Huang,
and H. Vincent Poor\vspace{-0pt}\\
  
\IEEEcompsocitemizethanks{ 
		\IEEEcompsocthanksitem
		
M. Zhang and J. Huang are with Department of
Information Engineering, The Chinese University of Hong Kong, Shatin, NT,
Hong Kong, E-mail: \{zm015, jwhuang\}@ie.cuhk.edu.hk;	J. Huang is also with School of Engineering and Science, The Chinese University of Hong Kong, Shenzhen; A. Arafa and H. V. Poor are with the Department of Electrical Engineering, Princeton University, NJ, USA, E-mail: \{aarafa, poor\}@princeton.edu. This work was supported in part by the Global Scholarship Programme for Research Excellence from CUHK, in part by the Overseas Research Attachment Programme from School of Engineering at CUHK, in part by the General Research Fund CUHK 14219016 from Hong Kong UGC, in part by the Presidential Fund from the Chinese University of Hong Kong, Shenzhen, and in part by the U.S. National Science Foundation under Grants CCF-0939370 and CCF-1513915.}
%
\vspace{-20pt}
}

\maketitle

\begin{abstract}
	We introduce the concept of a {\it fresh data market}, in which a destination user requests, and pays for, fresh data updates from a source provider. Data freshness is captured by the {\it age of information} (AoI) metric, defined as the time elapsed since the latest update has reached the destination. The source incurs an operational cost, modeled as an increasing convex function of the number of updates. The destination incurs an age-related cost, modeled as an increasing convex function of the AoI. The source charges the destination for each update and designs a pricing mechanism to maximize its profit; the destination on the other hand chooses a data update schedule to minimize the summation of its payments to the source and its age-related cost. The interaction among the source and destination is hence game-theoretic. Motivated by the existing pricing literature, we first study a {\it time-dependent} pricing scheme, in which the price for each update depends on when it is requested. We show in this case that the game equilibrium leads to only one data update, which does not yield the maximum profit to the source. This motivates us to consider a {\it quantity-based} pricing scheme, in which the price of each update depends on how many updates have been previously requested. We show that among all pricing schemes in which the price of an update may vary according to both time and quantity, the quantity-based pricing scheme performs best: it maximizes the source's profit and minimizes the social cost of the system, defined as the aggregate source's operational cost and the destination's age-related cost. Numerical results show that the optimal  quantity-based pricing can be $27\%$ more profitable for the source and incurs $54\%$ less social cost, compared with the optimal  time-dependent pricing.
\end{abstract}

\section{Introduction}

\subsection{Motivation}

Information usually has the greatest value when it is fresh \cite[p. 56]{fresh}. Data freshness is becoming increasingly significant due to the fast growth of the number of mobile devices and the dramatic increase of real-time applications: news updates, traffic alerts, stock quotes, and social media updates. In addition, timely information updates are also critical in real-time monitoring, data analytics, and control systems. For instance, real-time knowledge of traffic information and the speed of motor vehicles is crucial in autonomous driving and unmanned aerial vehicles. Another instance is for phasor data updates in power grid stabilization systems and application program interface (API) monitoring \cite{api}. Examples of real-time datasets include real-time map data and traffic data, such as the Google Maps Platform \cite{map}. A suitable candidate metric to measure the freshness of data is the {\it age of information} (AoI) metric, introduced in \cite{AoI1,AoI20}, which measures the amount of time elapsed since the most recent data update. 

\begin{figure}
	\begin{centering}
		\includegraphics[scale=0.35]{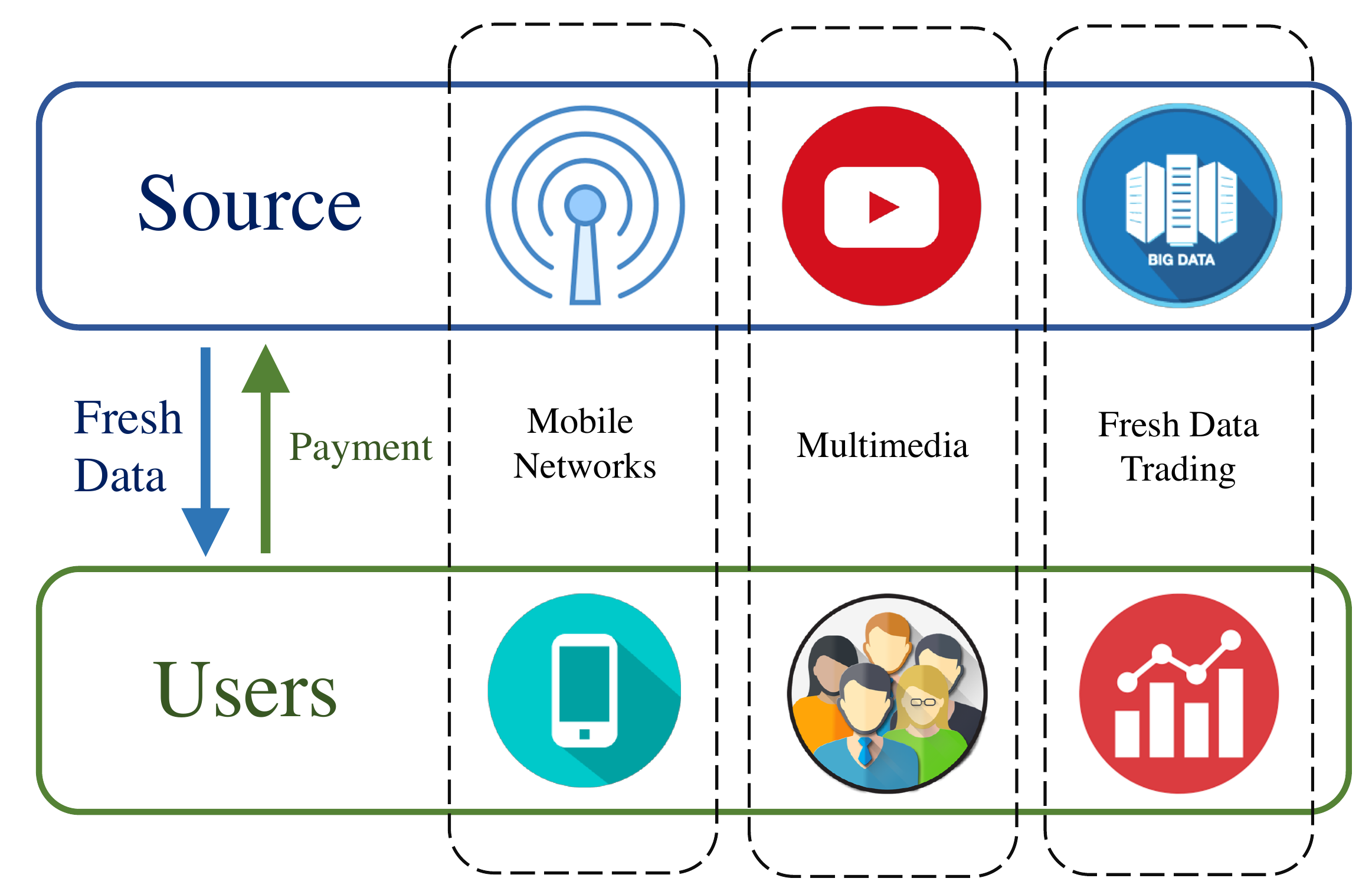}
		
		\vspace{-0.25cm}
		\caption{Examples of potential fresh data markets.}
		\label{System}
		
		\vspace{-0.5cm}
	\end{centering}
\end{figure}

However, the availability of fresh data relies on frequent data generation, processing, and transmission, which can lead to significant operational costs for the data provider. Such operational costs make the pricing design of an essential role in the data market, as pricing provides an incentive for the data provider to update the data and prohibits the data users (receivers) from requesting data updates unnecessarily often. This is quite different from the traditional study of pricing in networks, which often aims at maximizing a network operator's revenue and control the network congestion level. The pricing for fresh data is under-explored, as all existing pricing schemes for communication systems assume that a consumer's satisfaction with the service depends mainly on the quantity/quality of the service received without considering its timeliness.  
Such an interaction between data providers and users requesting fresh data leads to the
{\it fresh data market}, examples of which are shown in Fig. \ref{System}.
This paper tries to partially fill in the gap by considering a single source-destination pair, and addressing the following key question:
\begin{question}
	How should the source choose the pricing scheme to maximize its profit in a fresh data market? 
\end{question}

\subsection{Solution Approach and Contributions}

As the first step toward studying the pricing mechanism design for fresh data, we consider two types of pricing schemes. The first one is a \textit{time-dependent} pricing scheme, in which the  source of fresh data  prices each data update based on the time at which the update is requested. Due to the nature of the AoI, the destination's desire for updates increases as time (since the most recent update) goes by, which makes it potentially profitable to explore this time sensitivity. This pricing scheme is also motivated by many existing time-dependent pricing schemes  (in which users are not age-sensitive) of mobile networks, e.g., \cite{Pricing1,Pricing2,Pricing3,Pricing4,time2,time3,time4,time5,time6}.

The second pricing scheme that we consider is a \textit{quantity-based} pricing scheme, in which the price for each update depends on the number of updates requested so far (but does not depend on the timing of the updates). Such a pricing scheme is also known as \textit{second-degree price discrimination} or \textit{volume discount} \cite{Price}, and is motivated by practical pricing schemes (e.g., for mobile data plans  and data analytics \cite{api}).

The challenge of designing a proper pricing scheme for fresh data is two-fold. First, different from the classical pricing setting, e.g., \cite{Pricing1,Pricing2,Pricing3,Pricing4,time2,time3,time4,time5,time6}, the demands for fresh data over time are interdependent due to the nature of AoI. That is, the desire for an update at each time instance depends on the time elapsed since the latest update. Hence, the source's pricing scheme choice needs to take such interdependence overall the entire period into consideration. 
Second, in the case of the time-dependent pricing scheme  design, one needs to optimize a continuous-time pricing function, i.e., solve an infinite dimensional optimization problem.
The above discussion motivates our consideration of the following question:
\begin{question}
	How profitable is it for the source to exploit the time sensitivity in designing the  pricing scheme for fresh data?
\end{question}
We summarize our approaches and contributions as follows:
\begin{itemize}
	\item \emph{Fresh Data Market Modeling. } To the best of our knowledge, this paper presents the first model of a fresh data market, in which an age-sensitive destination interacts with a source data provider.
	\item \emph{Time-Dependent Pricing Scheme.} We study a time-dependent pricing scheme for the fresh data market, aiming at exploiting the time sensitivity. We show that at the optimal (equilibrium) time-dependent pricing scheme, the source sends only one update, and hence exploiting time sensitivity may not enhance profitability.
	\item \emph{Quantity-Based Pricing Scheme.} We propose a quantity-based pricing scheme, and show that it is more profitable than the time-dependent pricing scheme. We further prove that it maximizes the profit among all classes of time-and-quantity dependent pricing schemes, and that it minimizes the social cost of the system: the sum of the source's operational cost and the destination's age-related cost.
	\item \emph{Simulation Results.} The numerical results show that the optimal quantity-based pricing scheme can be $27\%$ more profitable and incurs $54\%$ less social cost, compared with the optimal time-dependent pricing scheme on average.
\end{itemize}

We organize the rest of this paper as follows. In Section \ref{Relate}, we discuss some related work. In Section \ref{Sysm}, we describe the system model and the game-theoretic problem formulation. In Sections \ref{TimeDepen} and \ref{CountDepen}, we develop the time-dependent and the quantity-based pricing schemes, respectively. We then relate the two schemes and mention some relevant properties in Section \ref{Prop}. We provide some numerical results in Section \ref{Numerical} to evaluate the performance of the two pricing schemes, and conclude the paper in Section \ref{Conclusion}.

\section{Related Work}\label{Relate}

The concept age-of-information was first proposed as a metric of data freshness in the studies of databases\cite{AoI1,AoI20} in the 1990s. In recent years, there have been many excellent works focusing on the optimization of scheduling policies in terms of minimizing the AoI in various system settings, see, e.g., \cite{AoI2,AoI3,AoI4,AoI42,AoI5,AoI6,New1,New2,NewAge1,NewAge2,energy1,energy2,AoI7,AoI12,AoI13,AoI9,AoI10,AoI11}. In \cite{AoI2}, Kaul \textit{et al.} recognized the importance of real-time status updates in networks. In \cite{AoI4,AoI42}, He \textit{et al.} investigated the
NP-hardness of minimizing the AoI in  scheduling general wireless networks.
In \cite{AoI5}, Kadota \textit{et al.} studied the scheduling problem in a wireless network with a single base station and multiple destinations. In \cite{AoI6}, Kam \textit{et al.} investigated the AoI for a status updating system through a network cloud. In \cite{AoI7}, Sun \textit{et al.} studied the optimal management of the fresh information updates. 
References \cite{New1} and \cite{New2} studied the optimal wireless network scheduling with the interference constraint and the throughput constraint, respectively.
The AoI consideration has recently gained some attention in energy harvesting communication systems, e.g., \cite{AoI3,energy1,AoI12,energy2,AoI13} and Internet of Things systems, e.g., \cite{NewAge1,NewAge2}.
Several existing studies focused on game-theoretic interactions in interference channels,
without considering the interactions in a fresh data market or the pricing scheme design \cite{AoI9,AoI10,AoI11}.

There exists a rich literature on the pricing mechanism design and revenue management in communication networks (please refer to \cite{Pricing1,Pricing2,Pricing3,Pricing4,time2,time3,time4,time5,time6}, surveys in \cite{survey,time1}, and references therein). 
Specifically, 
time-dependent pricing has also been extensively studied, e.g., \cite{Pricing3,time2,time3,time4,time5,time6} while
a few works focused on the quantity-based pricing and other forms of price differentiation for the Internet service providers, e.g., \cite{Pricing1,Pricing2,Pricing4}. 
These works assumed that destinations are only interested in the throughput/rate received instead of the data freshness.

References \cite{AoIEco1,AoIEco2} are the most closely-related works to ours. In \cite{AoIEco1}, a repeated game is studied between two AoI-aware platforms, yet without studying pricing schemes. While in \cite{AoIEco2}, the authors considered a system in which the destination designs a dynamic pricing scheme to incentivize sensors to provide fresh updates, with random data arrivals. Different from\cite{AoIEco2}, our considered pricing schemes are designed by the source, which is motivated by most practical communication/data systems in which sources are price designers while the destinations are price takers.

\section{System Model}\label{Sysm}

\subsection{System Overview}

\subsubsection{Single-Source Single-destination System} 
We consider an information update system, in which one source node generates data packets and sends them to one destination through a channel.\footnote{We note that the single-source single-destination model has been widely considered in the AoI literature (e.g.,
	\cite{AoI3,AoI6,AoI7,AoI12,AoI13}). In addition, the insights derived from this model allow us to potentially extend the results to the multi-destination scenarios.}


\subsubsection{Data Updates  and Age of Information}  We consider a fixed time period of $\mathcal{T}=[0,T]$, during which the source sends its updates to the destination. We consider a generate-at-will model, as in, e.g., \cite{AoI3,energy1,AoI12,energy2,AoI13}, in which the source is able to generate and send a new update when requested by the destination. Updates reach the destination instantly, with negligible transmission time, as in, e.g., \cite{energy1,AoI12,energy2}.

We denote by $S_k\in\mathcal{T}$ the transmission time of the $k$-th update. The set of all update time instances is $\mathcal{S}\triangleq\{S_k\}$. Let $K$ denote the number of total updates, i.e., $|\mathcal{S}|=K$, where $|\cdot|$ denotes the cardinality of a set. The set $\mathcal{S}$ (and hence the value of $K$) is a decision variable of the destination. 

To measure the freshness of data, let us define the AoI $\Delta_t(\mathcal{S})$ at time $t$ as \cite{AoI1,AoI2}
\begin{align}
\Delta_{t}(\mathcal{S})=t-U_{t},
\end{align}
where $U_t$ is the time stamp of the most recently received update before time $t$, i.e., $U_{t}=\max_{S_{k}\leq t}\{S_{k}\}$.

\subsubsection{Source's Operational Cost and Pricing}  
We denote the source's operational cost by $C(K)$, which is modeled as an \textit{increasing convex} function in the number of updates $K$, with $C(0)=0$.\footnote{Formally, $C(\cdot)$ is defined on the set of non-negative real numbers $\mathbb{R}_+$, and then evaluated on the set of natural numbers $\mathbb{N}$.} This can represent transmission costs in case the source is a network operator, or the cost of generating, processing and transmission commission in case the source is a content/data provider.

The source designs the pricing scheme for sending the data updates. We consider a general scheme in which the price for a particular update may depend on the time of the update request and the number of previously requested updates. We denote by $p(t,k): \mathcal{T}\times\mathbb{N}\rightarrow \mathbb{R}_+$ the pricing function, with $p(t',k')$ being the price of the $k'$th update if requested at time $t'$. Note that we denote by $p(t,k)$ the function itself, while we denote by $p(t',k')$, i.e., using any argument other than $(t,k)$ specifically, the value of the function.
As mentioned, such a pricing scheme is motivated by (i) the time-sensitive  demand for an update due to the nature of AoI, and (ii) the wide consideration  of both time-dependent and quantity-based pricing schemes in practice. Under a pricing scheme $p(t,k)$, the destination's total payment to the source over the entire period is $
P(\mathcal{S})=\sum_{k=1}^{K}p(S_k,k)$.

\begin{figure}
	\begin{centering}
		\includegraphics[scale=0.27]{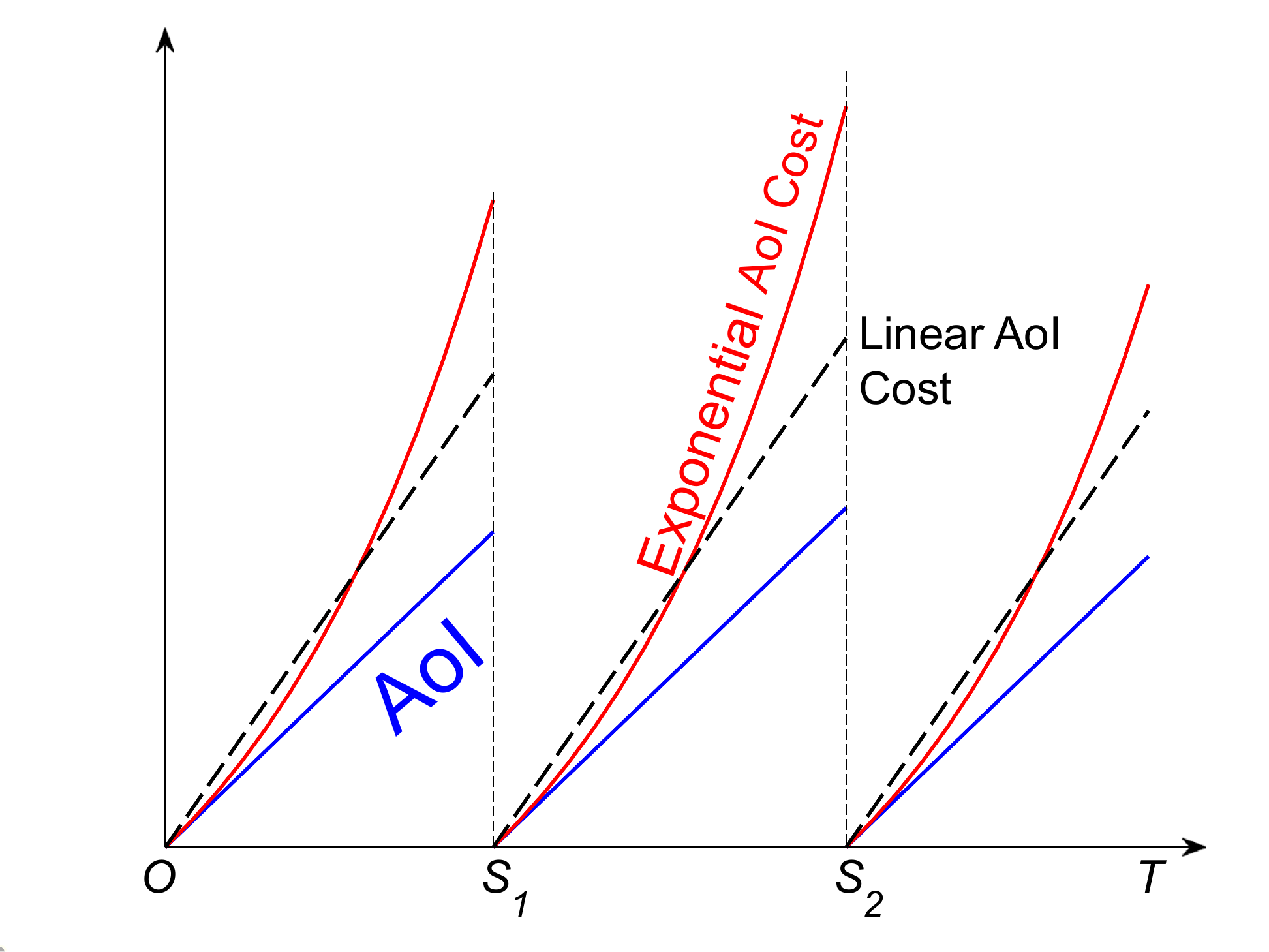}
		
		\vspace{-0.5cm}
		\caption{Illustrations of AoI $\Delta_t$ and two types of AoI costs $f(\Delta_t)$. There are two updates at $S_1$ and $S_2$.}
		\label{AoI}
	\end{centering}
	\vspace{-0.5cm}
\end{figure}

\subsubsection{Destination's AoI Cost} 
Besides the payment $P(\mathcal{S})$, the destination also experiences an AoI cost $f(\Delta_t)$ related to the destination's desire for the new data update.\footnote{As the first work considering the pricing scheme design for fresh data, we assume the benefit of receiving the data is constant, i.e. independent of the total number of updates. }
We assume that $f(\Delta_t)$ is \textit{increasing} and {\it convex} in $\Delta_t$.\footnote{An example of this AoI cost model exists in the online learning in real-time applications such as online advertisement placement and online Web ranking, in which fresh data is critical \cite{online1,online2,convex}.}
Let $\Gamma(\mathcal{S})$ denote the aggregate AoI cost over the entire period $\mathcal{T}$, defined as
\begin{align}
\Gamma(\mathcal{S})\triangleq\int_{0}^{T} f(\Delta_t(\mathcal{S}))dt.\label{gamma}
\end{align}
Fig. \ref{AoI} illustrates  the AoI, an exponential AoI cost, and a linear AoI cost.

\begin{figure}[t]
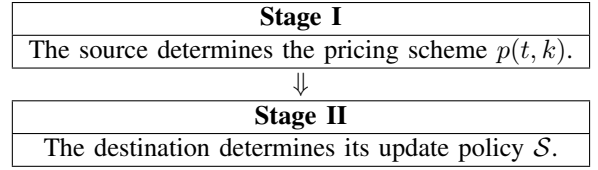

	\centering
	\begin{tabular}{c}
		\hline
		\multicolumn{1}{|c|}{ \textbf{Stage I}}        \\ \hline
		\multicolumn{1}{|c|}{The source determines the pricing scheme $p(t,k)$.} \\ \hline
		$\Downarrow   $                                  \\ \hline
		\multicolumn{1}{|c|}{\textbf{Stage II}}        \\ \hline
		\multicolumn{1}{|c|}{The destination determines its update policy $\mathcal{S}$.}         \\ \hline
	\end{tabular}
	\caption{Two-stage Stackelberg game.}	\label{Stack1}
	
	\vspace{-0.5cm}
\end{figure}

\subsection{Stackelberg Game}

We model the interaction between the source and the destination as a
two-stage Stackelberg game as shown in Fig. \ref{Stack1}. 
Specifically, in Stage I, the source determines the pricing scheme function $p(t,k)$ at the beginning of the period, in order to maximize its profit, given by the payment it receives minus its operational cost, as follows:
\vspace{-0.13cm}
\begin{subequations}\label{source}
	\begin{align}
	{\rm \mathbf{Source:}}~&\max_{p(t,k)}~P(\mathcal{S}^*(p(t,k)))-C(|\mathcal{S}^*(p(t,k))|),\\
	&~~{\rm s.t.}~~p(t',k')\geq0,~\forall t'\in\mathcal{T}, k'\in\mathbb{N},
	\end{align}
\end{subequations}
where $\mathcal{S}^*(p(t,k))$ is the destination's optimal update policy, in response to the pricing scheme chosen by the source, which is defined below.

In Stage II, the destination decides its update policy 	to minimize its overall cost (aggregate AoI cost plus payment):
\vspace{-0.06cm}
\begin{align}\label{destination}
{\rm \mathbf{Destination:}} ~\mathcal{S}^*(p(t,k))=\arg\min_{\mathcal{S}\in\Phi}~\Gamma(\mathcal{S})+P(\mathcal{S}),
\end{align}
where $\Phi$ is  the set of all feasible $\mathcal{S}$, given by $\Phi=\cup_{K'\in\mathbb{N}}\Phi^{K'}$ and $\Phi^{K'}$  is the set of all transmission times $\mathcal{S}=\{S_1,S_2,...,S_{K'}\}$, with $S_k'\in\mathcal{T}$ and $S_k\geq S_{k-1}$ for all $1\leq k\leq K'$.
%
%

In the following  two sections, we will separately consider two special cases of $p(t,k)$: $p(t)$ and $p_q(k)$.
We note that analyzing the simplified pricing schemes is still challenging. First, the optimal pure time-dependent pricing scheme involves solving an infinite-dimensional optimization problem. Second, the pricing scheme needs to take the optimal decisions $\mathcal{S}^*(\cdot)$ over the whole period into consideration.

\section{Time-Dependent Pricing Scheme}\label{TimeDepen}
In this section, we consider a (pure) \textit{time-dependent} pricing scheme, in which the price function only depends on the time at which the update is requested and does not depend on the number of updates. 

We derive the \textit{(Stackelberg)} equilibrium price-update profile  $({p}^*(t),\mathcal{S}^*(p^*(t)))$ using  backward induction. First,  given any pricing scheme $p(t)$  in Stage I, we characterize the destination's update policy $\mathcal{S}^*(p(t))$ that minimizes its overall cost in Stage II. Then in Stage I, by characterizing the equilibrium pricing structure, we convert the continuous pricing function optimization into a vector one, based on which
we characterize the source's optimal pricing scheme $p^*(t)$.

\subsection{Destination's Update Policy in Stage II}

Recall that   $K$ is  the total number of updates. Let $x_k$ denote the $k$th interarrival  time, which is the time elapsed between the generation of ($k-1$)-th update and  $k$-th update, i.e., $x_k$ is 
\vspace{-0.35cm}
\begin{align}
x_k\triangleq S_k-S_{k-1},~\forall k\in \mathcal{K}(K+1),
\end{align}
where $\mathcal{K}(K+1)=\{1,2,...,K+1\}$, $S_0=0$, and $S_{K+1}=T$.

%

To analyze the aggregate AoI cost function $\Gamma(\mathcal{S})$ in \eqref{gamma}, we define \vspace{-0.4cm}
\begin{align}
F(x)\triangleq\int_0^x f(t)dt,
\end{align}
based on which we have $\Gamma(\mathcal{S})=\sum_{k\in\mathcal{K}(K+1)}F(x_k)$.

Given the pricing scheme $p(t)$, the destination's problem in \eqref{destination} is equivalent to
\vspace{-0.3cm}
\begin{subequations}
	\begin{align}
	&\min_{K\in\mathbb{N}\cup\{0\},\bs{x}\in\mathbb{R}_{++}^{K+1}}~\sum_{k=1}^{K+1}F(x_k)+\sum_{k=1}^Kp\left(\sum_{j\leq k}x_j\right),\\
	&~~~~~~~~~{\rm s.t.}~~~~~~~~~\sum_{k=1}^{K+1}x_k=T,
	\end{align}
\end{subequations}
where $\bs{x}=\{x_k\}_{k\in\mathcal{K}(K+1)}$ and $\mathbb{R}_{++}^{K+1}$ is the space of $(K+1)$-dimensional positive vectors (i.e., the value of every entry is positive).


%

%

\begin{figure}
	\begin{centering}
		\includegraphics[scale=0.2]{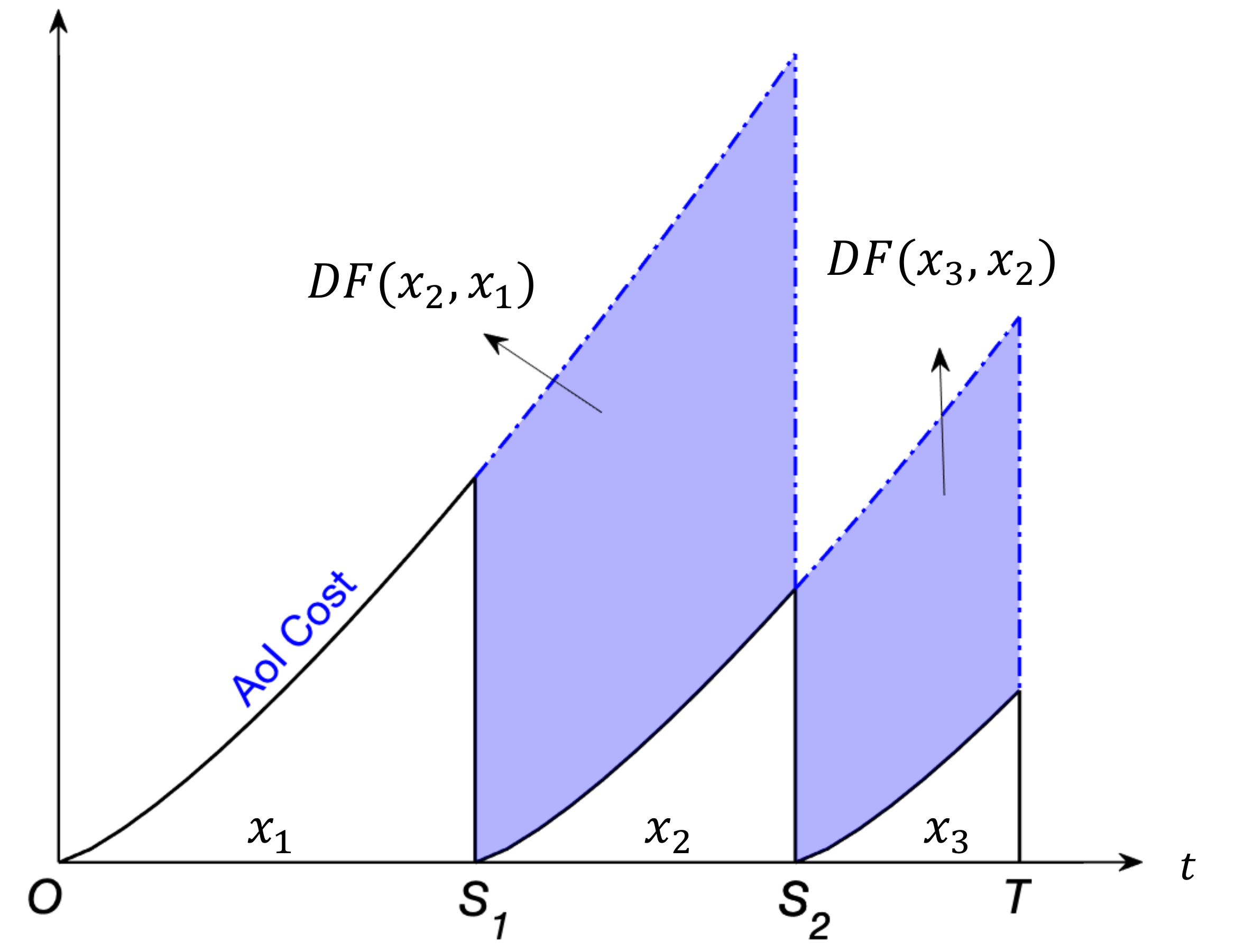}
		\vspace{-0.35cm}
		\caption{An illustrative example of the differential aggregate AoI cost function and Lemma \ref{L2}. }
		\label{Lemma2}
	\end{centering}
	\vspace{-0.4cm}
\end{figure}

To understand when the destination would choose to update, we define the differential aggregate AoI cost function as
\vspace{-0.1cm}
\begin{align}
DF(x,y)\triangleq \int_0^x [f(t+y)-f(t)]dt.
\end{align}
As illustrated in  Fig. \ref{Lemma2}, for each update $k$,  $DF(x_{k+1},x_k)$ is the aggregate AoI cost increase if the destination changes its update policy from $\mathcal{S}$ to  $\mathcal{S}\backslash\{S_k\}$ (removes the update at  $S_k$).
We  now introduce the following lemma: 
\begin{lemma}\label{L2}
	Any equilibrium price-update tuple $(p^*(t),K^*,\bs{x}^*)$  should satisfy 
	\vspace{-0.2cm}
	\begin{align}
	p^*\left(\sum_{j=1}^kx_j^*\right)=
	DF(x_{k+1}^*,x_k^*),~&~\forall k\in\mathcal{K}(K^*+1).\label{price}
	\end{align}
\end{lemma}

\rev{\noindent \textit{Proof Sketch}: For each $k$-th update, the differential aggregate AoI cost equals  the destination's maximal willingness to pay. Hence, if $p^*\left(\sum_{j=1}^kx_j^*\right)>
DF(x_{k+1}^*,x_k^*)$, then the destination would prefer not to update at $S_k$, contradicting to the fact that  $(p^*(t),K^*,\bs{x}^*)$ is an equilibrium.
In addition, if $p^*\left(\sum_{j=1}^kx_j^*\right)<DF(x_{k+1}^*,x_k^*)$, we can show that the source can always properly increase $p^*\left(\sum_{j=1}^kx_j^*\right)$. The increase in the price does not change the destination's optimal solution $(K^*,\bs{x}^*)$, and hence increases the source's profit. This contradicts the fact that  $(p^*(t),K^*,\bs{x}^*)$ is an equilibrium.
\hfill\qedsymbol}

Note that given that the optimal pricing scheme satisfies \eqref{price}, there might exist multiple optimal update policies as the
solutions of problem \eqref{destination}. This may lead to a multi-valued source's profit and thus an ill-defined  problem \eqref{source}. To ensure the uniqueness of the received profit for the source,  
 one can impose infinitely large prices to
ensure that  the destination does not update at any time instance other than $\sum_{j=1}^kx_j^*$ for all $k\in\mathcal{K}(K^*+1)$.

\subsection{Source's Time-Dependent Pricing Design in Stage I}
Based on Lemma \ref{L2},  we can reformulate the time-dependent pricing scheme as follows (the proof is omitted due to space limits).
\begin{theorem}\label{L44}
	The time-dependent pricing problem  in \eqref{source} is equivalent to the following problem:
	\vspace{-0.2cm}
	\begin{subequations}\label{Refor}
		\begin{align}
		\max_{K\in\mathbb{N}\cup\{0\},\bs{x}\in\mathbb{R}^{K+1}_{++}}&~~\sum_{k=1}^{K}DF(x_{k+1},x_k)-C(K),\label{Refor1}\\
		{\rm s.t.}~~~~~~~&~~\sum_{k=1}^{K+1}x_k=T\label{dsa}.
		\end{align}
	\end{subequations}
\end{theorem}

The decision variables in problem \eqref{Refor} correspond to the interarrival time interval vector $\bs{x}$ instead of the continuous-time pricing function $p(t)$. 
By converting a continuous function optimization problem into a vector optimization problem, we significantly simplify the problem.
We are now ready to present the following result:
\begin{theorem}\label{T1}
	There will be only one update (i.e., $K^*=1$) under any equilibrium time-dependent pricing scheme. 
\end{theorem}
One can prove Theorem \ref{T1}  by induction, showing that for an arbitrary time-dependent pricing scheme yielding more than $K>1$ updates ($K$-update pricing), there  always exists a pricing scheme leading to a single-update equilibrium that is more profitable. 
The following example illustrates this with a linear AoI cost function:


\begin{figure}
	\begin{centering}
		\includegraphics[scale=0.32]{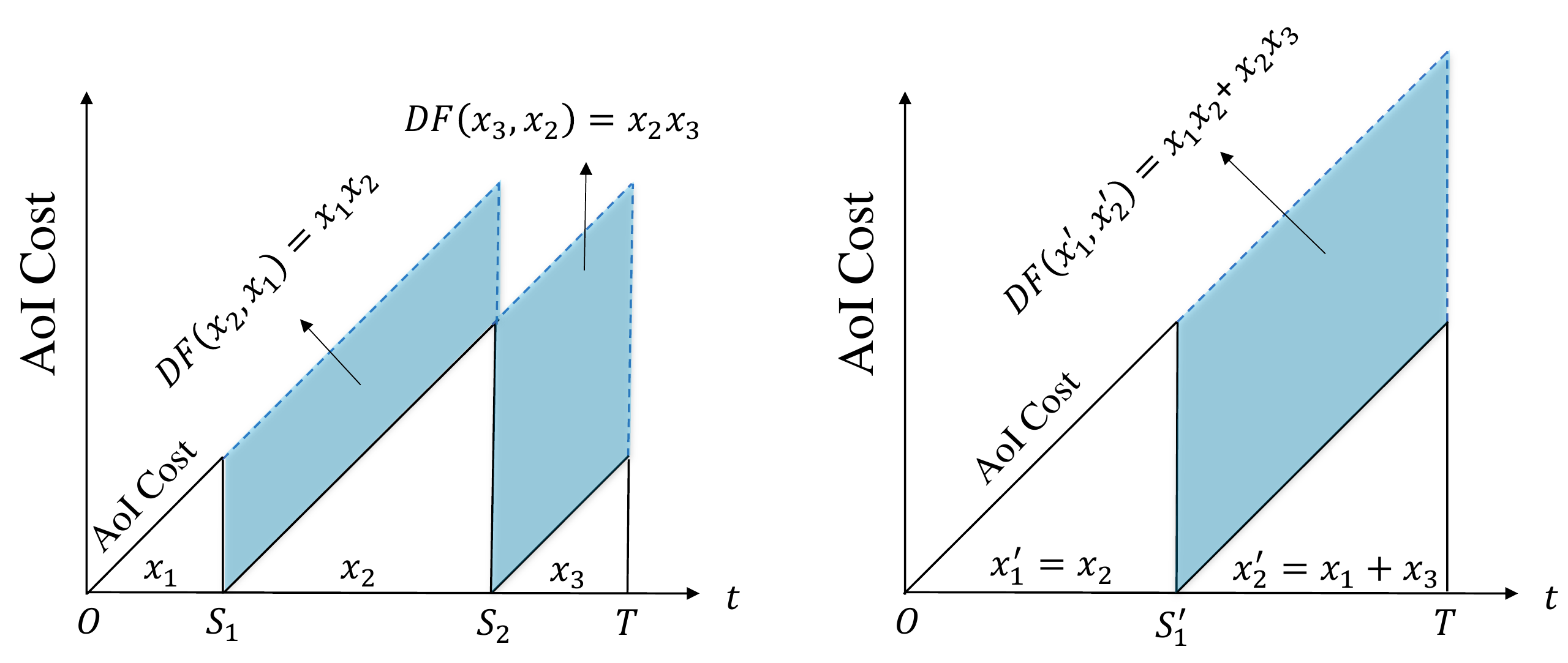}
		\vspace{-0.4cm}
		\caption{Illustrations of Example \ref{E1} with a linear cost function. Combining the first interval into the third interval maintains the payment.}
		\label{example1}
	\end{centering}
	\vspace{-0.4cm}
\end{figure}

\rev{\begin{example}\label{E1}
	Consider a linear AoI cost $f(\Delta_t)=\Delta_t$ and an arbitrary update policy $(K,\bs{x})$, as shown in Fig. \ref{example1}.
	
	\begin{itemize}
		\item \textbf{Base case:} When there are $K=2$ updates, as shown in Fig. \ref{example1}, the source's profit (the objective value in \eqref{Refor1}) is
		$x_1x_2+x_2x_3-C(2).$
		Consider another update policy $(K'=1,x_1',x_2')$ where $x_1'=x_2$ and $x_2'= x_3+x_1$. The objective value  in \eqref{Refor1} becomes
		$x_2(x_1+x_3)-C(1).$
		Comparing these two values, 
		we see that $(K',x_1',x_2')$ is strictly more profitable than $(K,\bs{x})$.
		\item  \textbf{Induction step:} Let $K\geq n$ and suppose the statement that, for an arbitrary $K$-update pricing, there exists a more profitable $(K-1)$-update pricing
		is true for $K=n$. The objective value in \eqref{Refor1} is
$\sum_{k=1}^Kx_kx_{k+1}-C(K).$
		Consider another update policy $(K'=K-1,\bs{x}')$ where $x_1'=x_2$, $x_2'= x_3+x_1$, and $x_k'=x_{k+1}$ for all other $k$. The objective value in \eqref{Refor1} becomes $
		(x_1+x_3)(x_2+x_4)+\sum_{k=4}^Kx_kx_{k+1}-C(K-1),$
		which is strictly larger than $P$. It is then readily verified that $(K',\bs{x}')$ is strictly more profitable than $(K,\bs{x})$. Based on induction, we can show that we can find a $(K'-1)$-update policy would be more profitable than the $(K',\bs{x}')$ policy. This eventually leads to the conclusion that a single update policy is the most profitable. 
	\end{itemize}
\end{example}}
Based on the above technique,  we can show that the above argument works for any increasing convex AoI cost function. The complete proof is omitted due to space limits.

To rule out trivial cases in which there is no update at the equilibrium, we adopt the following assumption:
\begin{assumption}\label{A2}
	The source's operational cost function $C(K)$ satisfies $C(1)\leq DF(T/2,T/2).$
\end{assumption}
Assumption \ref{A2} ensures that the operational cost for one update $C(1)$ is not larger than the maximal revenue for one update $DF(T/2,T/2)$,
as shown in the following optimal time-dependent pricing scheme:
\begin{proposition}\label{P11}
	There exists an optimal time-dependent pricing scheme such that\footnote{There exist multiple optimal pricing schemes; the only difference among all optimal pricing schemes are the prices for time instances other than $T/2$, which can be arbitrarily larger than $DF(T/2,T/2)$.}
	\begin{align}
	p^*({t})=DF\left(T/2,T/2\right),~\forall t\in\mathcal{T},
	\end{align}
	where the equilibrium update takes place at $S_1^*=T/2$.
\end{proposition}

Proposition \ref{P11} suggests the existence of an optimal time-dependent pricing scheme that is in fact \textit{time-invariant}. That is, although our original intention is to exploit the time sensitivity/flexibility of the destination through the time-dependent pricing, it turns out not to be very effective. This motivates us to consider a quantity-based pricing scheme next.

\section{Quantity-Based Pricing Scheme}\label{CountDepen}


In this section, we focus on a (pure) \textit{quantity-based} pricing scheme, in which the price for each update depends on the number of updates that  the destination has requested  so far.

The source determines the quantity-based pricing scheme 
$p_q(k)$ in Stage I, in which $p_q(k)$ represents the price for the $k$th update.
The payment from the destination will be 
$P_q(K)=\sum_{k=1}^{K}p_q(k)$.
Based on $\bs{p}_q=\{p_q(k)\}_{k\in\mathbb{N}}$,
the destination in Stage II chooses its update policy $(K,\bs{x})$.

We derive the \textit{(Stackelberg)} price-update equilibrium using the  bilevel optimization framework \cite{bilevel}. Specifically,
the bilevel optimization problem embeds the optimality condition of the low-level problem (the destination's problem \eqref{D}) into the  upper-level problem (the source's problem \eqref{source}).
We first characterize the  conditions of the destination's update policy $(K^*(\bs{p}_q),\bs{x}^*(\bs{p}_q))$ that minimize its overall cost in Stage II. 
We then substitute such conditions into the constraint set of the source's pricing problem in Stage I in order to
characterize the source's optimal pricing $\bs{p}_q^*$ accordingly. We use  $(K^*,\bs{x}^*)$ to denote the equilibrium update policy, i.e., $(K^*,\bs{x}^*)=(K^*(\bs{p}_q^*),\bs{x}^*(\bs{p}_q^*))$.

\subsection{Destination's Update Policy in Stage II} Given the quantity-based pricing scheme $\bs{p}_q$, the destination solves the following overall cost minimization problem:
\begin{subequations}\label{D}
	\begin{align}
	\min_{K\in\mathbb{N}\cup\{0\}, \bs{x}\in\mathbb{R}^{K+1}_{++}}~&\sum_{k=1}^{K+1} F(x_k)+\sum_{k=1}^Kp_q(k),\\
	{\rm s.t.}~~~~~~~~~&\sum_{k=1}^{K+1}x_k=T.\label{D2}
	\end{align}
\end{subequations}
If we fix the value of $K$ in \eqref{D}, then problem \eqref{D} is convex with respect to $\bs{x}$. The convexity allows us to 
exploit the Karush-Kuhn-Tucker (KKT) conditions on $\boldsymbol{x}$ to derive the following lemma (the proof is omitted due to space limits):
\begin{lemma}\label{L4}
	Under any given  quantity-based pricing scheme $\bs{p}_q$ in Stage I, the destination's optimal update policy $(K^*(\bs{p}_q),\bs{x}^*(\bs{p}_q))$ satisfies
	\vspace{-0.15cm}
	\begin{align}
	x_k^*(\bs{p}_q)=\frac{T}{K^*(\bs{p}_q)+1},~~\forall k\in\mathcal{K}(K^*(\bs{p}_q)+1). \label{L4eq}
	\end{align}
\end{lemma}

\begin{figure}
	\begin{centering}
		\includegraphics[scale=0.4]{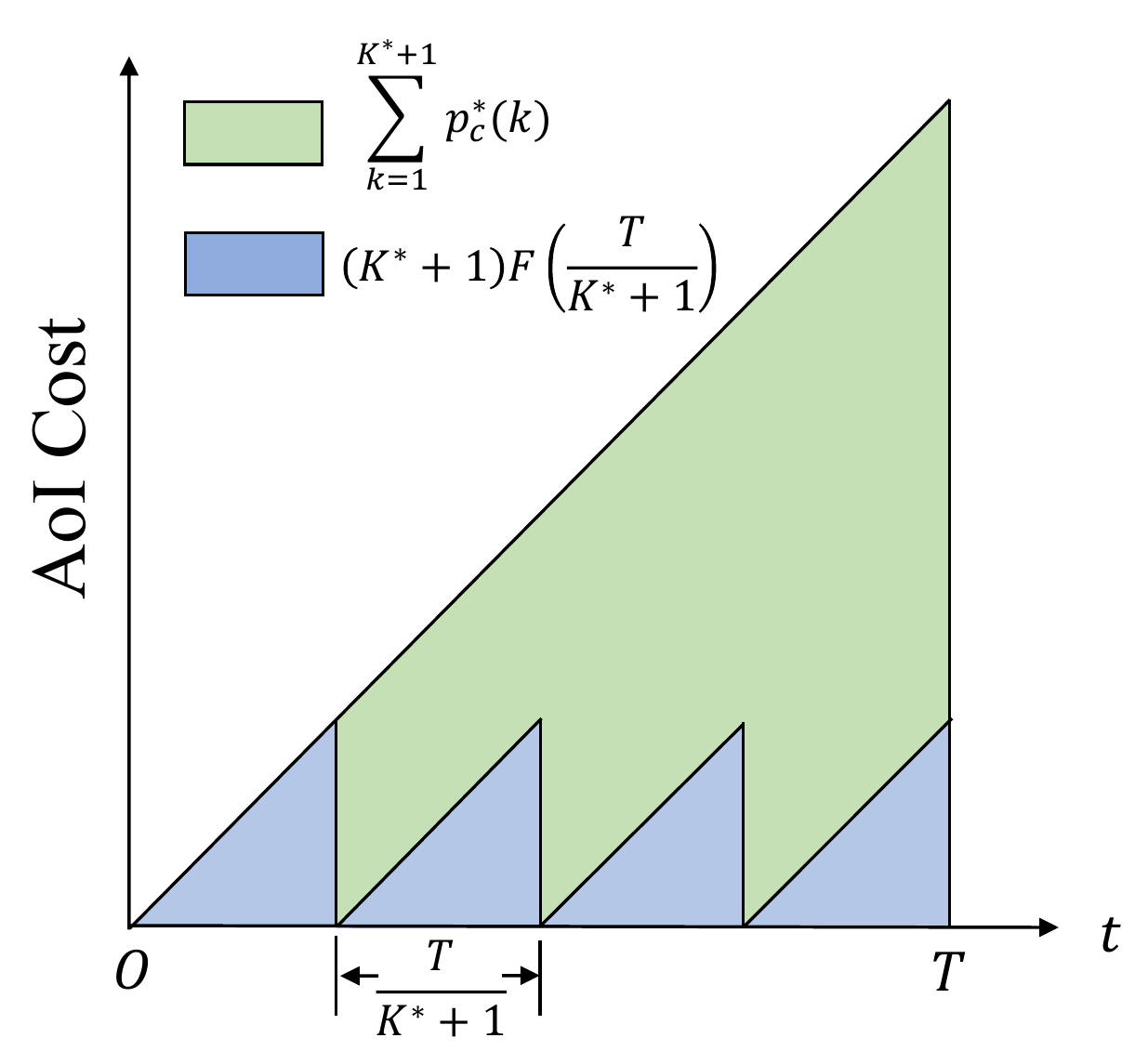}
		\vspace{-0.3cm}
		\caption{An illustrative example of Proposition \ref{T5} with an linear AoI cost function $f(\Delta_t)=\Delta_t$ and $K^*=3$. }
		\label{Theorem2}
	\end{centering}
	\vspace{-0.4cm}
\end{figure}

%
%

Lemma \ref{L4} indicates that the optimal update policy for the destination equalizes the inter-update time intervals. Hence, once the optimal interarrival time intervals is set according to \eqref{L4eq}, the destination would search for the optimal $K^*(\bs{p}_q)$ to minimize the objective in \eqref{D}:
\begin{align}
K^*(\bs{p}_q)\in\arg\min_{K'\in\mathbb{N}\cup\{0\}}\Upsilon(K',\bs{p}_q),\label{solution2}
\end{align}
where $\Upsilon(K',\bs{p}_q)$ is the overall cost given the equalized interarrival time intervals:
\begin{align}
\Upsilon(K',\bs{p}_q)\triangleq(K'+1)F\left(\frac{T}{K'+1}\right)+\sum_{k=1}^{K'}p_q(k).
\end{align}
\subsection{Source's Quantity-Based Pricing in Stage I}
Instead of solving both $K^*(\boldsymbol{p}_q)$ and $\boldsymbol{x}^*(\boldsymbol{p}_q)$ explicitly  in Stage II, we  apply the bilevel optimization  to solving the optimal quantity-based pricing $\bs{p}_q^*$ in Stage I. Doing so would lead to the price-update equilibrium of our entire two-stage game \cite{bilevel}.
By substituting the solutions \eqref{L4eq}-\eqref{solution2} into the source's pricing in \eqref{source}, we obtain the following bilevel problem:
\vspace{-0.2cm}
\begin{subequations}\label{Bilevel}
	\begin{align}
	{\rm \mathbf{Bilevel}:}~~\max_{\bs{p}_q, K,\bs{x}}&~~\sum_{k=1}^K p_q(k)-C(K),\\
	{\rm s.t.}&~~x_k=\frac{T}{K+1},~\forall k\in\mathcal{K}(K+1),\\
	&~~K\in\arg\min_{K'\in\mathbb{N}\cup\{0\}}\Upsilon(K',\bs{p}_q).\label{C3}
	\end{align}
\end{subequations}
In problem \eqref{Bilevel}, we treat $(K,\bs{x})$ as variables with the destination's behavior being part of the source's constraints. The optimal solution to the bilevel optimization problem \eqref{Bilevel} is exactly the equilibrium $(\bs{p}_q^*,K^*,\bs{x}^*)$ \cite{bilevel}.

The bilevel optimization in \eqref{Bilevel} leads to the following result:
\begin{proposition}\label{T5}
	The equilibrium update count $K^*$ and the optimal  quantity-based pricing scheme $\bs{p}_q^*$ satisfy
	\vspace{-0.25cm}
	\begin{align}
	\hspace{-2cm}	\sum_{k=1}^{K^*}p_q^*(k)&=F(T)-(K^*+1)F\left(\frac{T}{K^*+1}\right),\label{Ds}\\
	\sum_{k=1}^{K'}p_q^*(k)&\geq F(T)-(K'+1)F\!\left(\frac{T}{K'+1}\right),\forall K'\!\in\mathbb{N}\backslash\!\{\!K^*\!\}.\label{Da}
	\end{align}
	\vspace{-0.45cm}
\end{proposition}

\rev{\noindent \textit{Proof Sketch}: 
	 Fig. \ref{Theorem2} provides an illustrative example to  understand Proposition \ref{T5}. The area of the blue region is the aggregate AoI cost of the optimal updates, $(K^*+1)F(T/(K^*+1))$; the area of the blue region plus the green region is the aggregate AoI cost of a no-update scheme $F(T)$. The area of the green region
	 	$F(T)-(K^*+1)F \left(T/(K^*+1)\right)$ is the aggregate AoI cost difference between these two schemes.
	 
We prove that inequality \eqref{Da} together with \eqref{Ds} will  ensure that constraint \eqref{C3} holds. Specifically, if \eqref{Da} is not satisfied
or if $\sum_{k=1}^{K^*}p_q^*(k)>F(T)-(K^*+1)F\left(T/(K^*+1)\right)$,
then $K^*$
would violate constraint \eqref{C3}.
%
If $\sum_{k=1}^{K^*}p_q^*(k)<F(T)-(K^*+1)F\left(T/(K^*+1)\right)$, then the source can always properly increase $p_q^*(1)$ until \eqref{Ds} is satisfied. Such an increase does not violate constraint \eqref{C3} 
but improves the source's profit, contradicting to the optimality of $\bs{p}_q^*$.
\hfill\qedsymbol}

%
%

Substituting the pricing structure in \eqref{Ds} into \eqref{Bilevel}, we can obtain  $K^*$ through solving the following problem:\footnote{Note that $F(T)$ in \eqref{Ds} is a constant and hence is not considered in \eqref{PBK}.}
\begin{align}
\max_{K\in\mathbb{N}\cup\{0\}}~-(K+1)F\left(\frac{T}{K+1}\right)-C(K).\label{PBK}
\end{align}
To solve problem \eqref{PBK}, we first relax the constraint $K\in\mathbb{N}\cup\{0\}$ into $K\in\mathbb{R}_+$, hence transforming the integer programing problem \eqref{PBK} into a continuous optimization problem as follows:
\begin{align}
\max_{K\in\mathbb{R}_+}~-(K+1)F\left(\frac{T}{K+1}\right)-C(K),\label{PB2}
\end{align}
which is a convex problem.\footnote{To see the convexity of $(K+1)F\left(T/(K+1)\right)$, note that $(K+1)F\left(T/(K+1)\right)$ is the \emph{perspective} of function $F(T)$. The perspective of $F(T)$ is convex since $F(T)$ is convex \cite{convex}.}
We take the derivative of objective in \eqref{PB2} and obtain
\begin{align}
\underbrace{f\left(\frac{T}{K+1}\right)\frac{T}{K+1}-F\left(\frac{T}{K+1}\right)}_{\rm Marginal~Revenue}-\underbrace{C'(K)}_{\rm Marginal~Cost}.\label{s}
\end{align}
We can interpret the first term as the source's \textit{marginal revenue} in $K$ and the second term as the source's \textit{marginal cost} in $K$.

Based on the marginal revenue and the marginal cost in \eqref{s},
we define  a threshold update count $\hat{K}$ satisfying
\begin{subequations}\label{hatK}
	\begin{align}
	f\left(\frac{T}{\hat{K}+1}\right)\frac{T}{\hat{K}+1}-F\left(\frac{T}{\hat{K}+1}\right)&\geq C'(\hat{K}),\\
	f\left(\frac{T}{\hat{K}+2}\right)\frac{T}{\hat{K}+2}-F\left(\frac{T}{\hat{K}+2}\right)&< C'(\hat{K}+1).
	\end{align}
\end{subequations}
The threshold count $\hat{K}$ serves as one of the candidates for the optimal update count to  problem  \eqref{PBK} as shown next.\footnote{ Assumption \ref{A2} leads to the existence of a unique $\hat{K}$ satisfying \eqref{hatK}.}

\begin{proposition}\label{P3}
	The optimal  update count $K^*$ to problem in  \eqref{Bilevel} satisfies
	\vspace{-0.2cm}	
	\begin{align}\label{eq28}
	K^*=	\arg\min_{K\in\{\hat{K},\hat{K}+1\}}(K+1)F\left(\frac{T}{K+1}\right)+C(K).
	\end{align} 
\end{proposition}
\rev{\begin{IEEEproof}
	Let $K^*$ be the optimal solution to problem \eqref{PB2}. 
	By the definition of $\hat{K}$ in \eqref{hatK}, we have $\hat{K}\leq K^*\leq \hat{K}+1$. The convexity of the objective in \eqref{PB2} implies that the objective of \eqref{PB2} (which is also the objective of problem \eqref{PBK}) is non-decreasing in $K$ for all $K\leq K^*$ and is non-increasing in $K$ for all $K\geq K^*$. This implies an optimal solution to problem \eqref{PBK} is either $\hat{K}$ or $\hat{K}+1$. 
\end{IEEEproof}}


\begin{figure}
	\begin{centering}
		\includegraphics[scale=0.4]{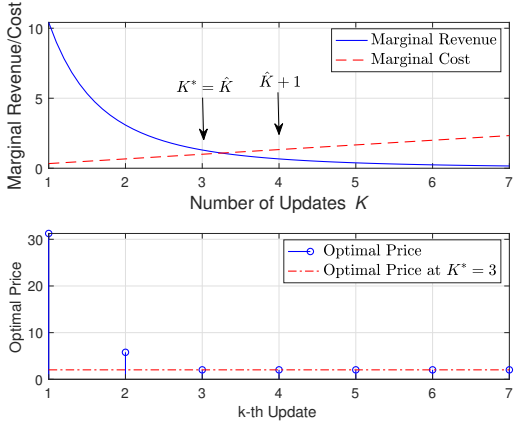}
		\vspace{-0.4cm}
		\caption{An illustrative example of the optimal  quantity-based pricing scheme in Proposition \ref{P4}. The destination's AoI cost is $f(\Delta_t)=\Delta_t^2$ and the source's operational cost is $C(K)=1/6K^2$.}
		\label{Fig6}
		\vspace{-0.4cm}
	\end{centering}
\end{figure}

After obtaining $K^*$, we can construct an optimal pricing scheme based on Proposition \ref{T5} as follows:

\begin{proposition}\label{P4}
	An optimal  quantity-based pricing $\bs{p}^*_q$ is
	\vspace{-0.2cm}
	\begin{align}\hspace{-0.5cm}\label{optimalpricing}
	&p^*_q(k)\\
	=&\begin{cases}
	0,~~~~~~~~~~~~~~~~~~~~~~~~~~~~~~~~~~~~~~~~~~~~{\rm if}~k=0,\\
	F(T)-(k+1)F(\frac{T}{k+1})-\sum_{j=1}^{k-1}p^*_q(j)+\epsilon,\\~~~~~~~~~~~~~~~~~~~~~~~~~~~~~~~~~~~~~~~~~~~~~~~{\rm if}~1< k< K^*,\\
	F(T)-(K^*+1)F(\frac{T}{K^*+1})-\sum_{j=1}^{K^*-1}p^*_q(j),~{\rm if}~k\geq K^*,
	\end{cases}\nonumber
	\end{align}
	where $\epsilon>0$ is an infinitesimal value to ensure \eqref{Da}.
\end{proposition}

Fig. \ref{Fig6} presents an illustrative example of \eqref{optimalpricing}. In  Fig. \ref{Fig6} (up), the marginal revenue intersects with the marginal cost in \eqref{s} at around $K=3.3$. Hence, the threshold update count is $\hat{K}=3$ based on \eqref{hatK}, and we can further verify based on \eqref{eq28} that the 
optimal update count is $K^*=3$. In Fig. \ref{Fig6} (down), we present the optimal quantity-based pricing scheme described in \eqref{optimalpricing}. As we can see,  the optimal price drops until the third update. The relatively high prices value of the first two update prices are to ensure \eqref{Da} holds for $K'=\{1,2\}$ while the relatively lower price starting from the third update is to ensure \eqref{Ds} holds.
%

\section{Properties}\label{Prop}
In this section, we study several properties of the pure quantity-based pricing and the pure time-dependent pricing.
Let $\Pi_q$ denote the achievable profit of the pure quantity-based pricing and $\Pi_t$  denote that of the pure time-dependent pricing.
We  first compare $\Pi_q$ with $\Pi_t$ in the following Proposition:
\begin{proposition}\label{P5}
	The achievable profit of the optimal quantity-based pricing $\Pi_q$ and that of the optimal  time-dependent pricing $\Pi_t$ satisfy
	\vspace{-0.3cm}
	\begin{align}
	\Pi_t\overset{(a)}{\leq} \Pi_q\overset{(b)}{<} 2\Pi_t.
	\end{align}
\end{proposition}
\vspace{-0.15cm}
\rev{\noindent \textit{Proof Sketch}: We can show that the time-dependent pricing scheme is a special case of the quantity-based pricing in Proposition \ref{P4} by fixing $K^*=1$, which proves (a). To prove (b), the destination's payment under the optimal time-dependent pricing scheme in  Proposition \ref{P11} is $DF(T/2,T/2)$, which we can show to be at least $F(T)/2$; while the destination's payment under the optimal quantity-based pricing is at most $F(T)$. 
\hfill\qedsymbol}

We are now ready to introduce the key result  of this paper:
\begin{theorem}[Profit Maximizing Structure]\label{P2}
	The optimal quantity-based pricing achieves the maximum source profit 
	among all possible time-and-quantity dependent pricing schemes in the form of  $p(t,k)$.
\end{theorem}
\rev{\noindent \textit{Proof Sketch}:
We first prove that, regardless of the pricing choice, the destination's payment is always upper-bounded by the AoI cost reduction as we discussed in Proposition \ref{P2}. Meanwhile, the optimal quantity-based pricing in \eqref{Da} attains such a bound. We then prove it is profit-maximizing.
\hfill\qedsymbol}

Theorem \ref{P2} implies that the relatively-simple quantity-based pricing scheme is already optimal. Hence, even without exploiting the time flexibility explicitly, it is still possible to obtain the optimal pricing structure, which again implies that utilizing time flexibility may not be necessary.

\begin{figure}[t]
	\begin{centering}
		
		\includegraphics[scale=0.28]{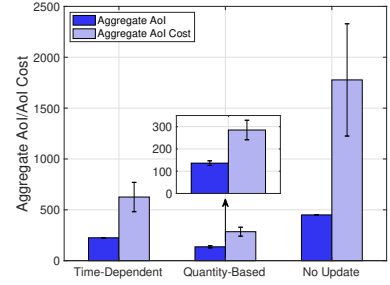}
		
		\vspace{-0.4cm}
		\caption{Performance comparison in terms of the aggregate AoI and the aggregate AoI cost. 
		}
		\label{Fig7}
		\vspace{-0.4cm}
	\end{centering}
\end{figure}

Next we introduce the social cost minimization problem, which  minimizes the sum of the destination's aggregate AoI cost and the source's operational cost:
\begin{subequations}\label{SCM}
\begin{align}
\hspace{-0.2cm}\min_{K\in\mathbb{N}\cup\{0\}, \bs{x}\in\mathbb{R}^{K+1}_{++}}&\sum_{k=1}^{K+1}F(x_k)+C(K),\\
~~~~~~~~{\rm s.t.}~~~~~~~&\sum_{k=1}^{K+1}x_k=T.
\end{align}
\end{subequations}
\begin{proposition}[Social Cost Minimization]\label{P6}
	The optimal quantity-based pricing scheme in \eqref{optimalpricing} leads to the optimal solution of \eqref{SCM}, i.e., minimizing the social cost. 
\end{proposition}
\rev{\noindent \textit{Proof Sketch}:
We first prove that the social cost minimizing update policy equalizes the time intervals, due to the similar reason as in Lemma \ref{L4}. In this case, problem \eqref{SCM} becomes equivalent to  problem \eqref{PBK}. Hence, the solution in \eqref{PBK} yields the minimal social cost.
\hfill\qedsymbol}


\section{Numerical Results}\label{Numerical}

In this section, we perform simulations to numerically compare both proposed pricing schemes regarding the aggregate AoI, the source's profit, and the social cost.

We consider a time interval of $T=30$ (days).\footnote{Examples of such a period of interest include the API monitoring platform \cite{api} and Google Map platform \cite{map}.}
The destination's AoI cost function is
$f(\Delta_t)=\Delta_t^\kappa,$
where  the exponent $\kappa$ is the \textit{destination's age sensitivity.}
Hence, the function $F(t)$ is
$F(t)=t^{\kappa+1}/(\kappa+1).$
The source has a cubic operational cost function, i.e., $C(K)=c \cdot K^3,$
where  $c$ is the source's \textit{operational cost coefficient}.
Let $\kappa$ follow a normal distribution $\mathcal{N}(1.5,0.2)$ truncated into the interval $[1,2]$; let $c$ follow a normal distribution $\mathcal{N}(6,1.5)$ truncated into the interval $[2,10]$.  




We compare the performance of three schemes: the optimal time-dependent pricing, the optimal quantity-based pricing, and a no-update benchmark. In Fig. \ref{Fig7},
we first compare the three schemes in terms of the aggregate AoI and the aggregate AoI cost. The no-update scheme incurs a much larger aggregate AoI than both proposed pricing schemes. Moreover, the optimal quantity-based pricing scheme incurs an  aggregate AoI  which is  only
$59\%$ of that incurred by the optimal time-dependent pricing. 
In terms of the aggregate AoI cost, we observe a similar trend. 

\begin{figure}[t]
	\begin{centering}
		\includegraphics[scale=0.28]{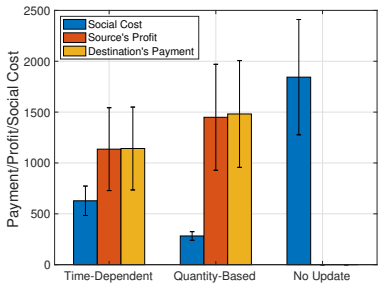}
		\vspace{-0.4cm}
		\caption{Performance comparison in terms of the social cost, the source's profit, and the destination's payment.
		}
		\label{Fig8}
		\vspace{-0.5cm}
	\end{centering}
\end{figure}

In Fig. \ref{Fig8}, we compare the three schemes in terms of the social cost, profit  (of the source), and payment (of the destination). First, we observe that  the optimal quantity-based pricing is $27\%$ more profitable than the optimal time-dependent pricing. Such an improvement is consistent with the analytical bounds in Proposition \ref{P5}. 
Finally, the optimal time-dependent pricing only incurs $34\%$ of the social cost of the no-update scheme, while the optimal quantity-based pricing further reduces the social cost and incurs only $46\%$ of that of the optimal time-dependent pricing. Note that the large standard deviations for both profits and payments of the proposed pricing schemes are mainly due to the large standard deviation of the aggregate AoI cost for the no-update scheme as shown in Fig. \ref{Fig7}.


\section{Conclusions}\label{Conclusion}

We have presented the first pricing scheme design for a fresh data market and proposed two pricing schemes. Our results have revealed that (i) the optimal time-dependent pricing scheme yields a single-update equilibrium, which does not effectively exploit the time flexibility, and 
(ii) the optimal quantity-based pricing scheme achieves the maximum profit for the source among all time-and-quantity dependent pricing schemes, and leads to the minimal social cost. Future work includes the extension to multi-destination scenarios, and studying incomplete user information settings.

\end{document}